\documentclass[conference]{IEEEtran}
\usepackage{cite}
\usepackage{amsmath,amssymb,amsfonts}
\usepackage{algorithmic}
\usepackage{graphicx}
\usepackage{textcomp}
\usepackage{hyperref} 
\usepackage{xcolor}
\usepackage{subcaption}
\usepackage{array}
\usepackage{array}
\usepackage{color, colortbl}
\hypersetup{
  pdftitle={},
  pdfauthor={},
    pdfnewwindow=true,      
    colorlinks=true,       
    linkcolor=blue,          
    citecolor=magenta,        
    filecolor=cyan,      
    urlcolor=teal
}

\definecolor{LightCyan}{rgb}{0.50,1,1}
\def\BibTeX{{\rm B\kern-.05em{\sc i\kern-.025em b}\kern-.08em
    T\kern-.1667em\lower.7ex\hbox{E}\kern-.125emX}}

\makeatletter
\newcommand{\linebreakand}{%
  \end{@IEEEauthorhalign}
  \hfill\mbox{}\par
  \mbox{}\hfill\begin{@IEEEauthorhalign}
}
\makeatother
\begin{document}

\IEEEpubid{\makebox[\columnwidth]{\textsuperscript{*}Manuscript accepted for IEEE EMBC 2024, Orlando, FL, USA. \hfill} \hspace{\columnsep}\makebox[\columnwidth]{ }}

\bstctlcite{IEEEexample:BSTcontrol} 

\title{Learning from Two Decades of Blood Pressure Data: Demography-Specific Patterns Across 75 Million Patient Encounters}
\author{\IEEEauthorblockN{Seyedeh~Somayyeh~Mousavi}
\IEEEauthorblockA{\textit{Department of Biomedical Informatics} \\
\textit{Emory University}\\
Atlanta, USA \\
\href{seyedeh.somayyeh.mousavi@emory.edu}{seyedeh.somayyeh.mousavi@emory.edu}}
\and
\IEEEauthorblockN{Yuting~Guo}
\IEEEauthorblockA{\textit{Department of Biomedical Informatics} \\
\textit{Emory University}\\
Atlanta, USA \\
\href{yuting.guo@emory.edu}{yuting.guo@emory.edu}
}
\and
\IEEEauthorblockN{Chad~Robichaux}
\IEEEauthorblockA{\textit{Department of Biomedical Informatics} \\
\textit{Emory University}\\
Atlanta, USA \\
\href{crobich@emory.edu}{crobich@emory.edu}}
\and
\linebreakand
\IEEEauthorblockN{Abeed~Sarker}
\IEEEauthorblockA{\textit{Department of Biomedical Informatics}\\ \textit{Emory University} \\
\textit{Department of Biomedical Engineering}\\ \textit{Georgia Institute of Technology}\\
Atlanta, USA \\
\href{abeed@dbmi.emory.edu}{abeed@dbmi.emory.edu}}
\and
\IEEEauthorblockN{Reza~Sameni}
\IEEEauthorblockA{\textit{Department of Biomedical Informatics}\\ \textit{Emory University} \\
\textit{Department of Biomedical Engineering}\\ \textit{Georgia Institute of Technology}\\
Atlanta, USA \\
\href{rsameni@dbmi.emory.edu}{rsameni@dbmi.emory.edu}}
}
\maketitle
\begin{abstract}
Hypertension is a global health concern with an increasing prevalence, underscoring the need for effective monitoring and analysis of blood pressure (BP) dynamics.

We analyzed a substantial BP dataset comprising 75,636,128 records from 2,054,462 unique patients collected between 2000 and 2022 at Emory Healthcare in Georgia, USA, representing a demographically diverse population.
We examined and compared population-wide statistics of bivariate changes in systolic BP (SBP) and diastolic BP (DBP) across sex, age, and race/ethnicity. The analysis revealed that males have higher BP levels than females and exhibit a distinct BP profile with age. Notably, average SBP consistently rises with age, whereas average DBP peaks in the forties age group. Among the ethnic groups studied, Blacks have marginally higher BPs and a greater standard deviation. We also discovered a significant correlation between SBP and DBP at the population level, a phenomenon not previously researched.

These results highlight the importance of demography-specific BP analysis for clinical diagnosis and provide valuable insights for developing personalized, demography-specific healthcare interventions.
\end{abstract}
 \begin{IEEEkeywords}
Blood Pressure; Demographics; Age; Race; Sex; Cuff-based Blood Pressure; Bias in Blood Pressure 
\end{IEEEkeywords}
\section{Introduction}
\label{sec:intro}
In 2021, the World Health Organization highlighted a concerning statistic: cardiovascular diseases (CVDs) are responsible for 32\% of all global deaths \cite{who}. Hypertension, often termed the ``silent killer'', is the most significant risk factor for strokes and heart attacks \cite{kalehoff2020story}.
Blood pressure (BP) monitoring is critical in diagnosing and mitigating the risk of CVDs \cite{Mousavi2024-qg}.
There is an increasing shift towards ambulatory BP monitoring, which allows patients to monitor their BP outside clinical settings and more frequently \cite{Mousavi2024-qg}. This approach, while convenient, raises questions about its accuracy vs traditional in-clinic measurements. The existing benchmarks for delineating normal, borderline, and hypertensive BP categories were established several decades ago, forming the current standards of CVD risk evaluation \cite{Whelton2018}. Nevertheless, these guidelines were established on relatively small populations, lacking individualization for demographic factors like sex, age, and race. The benchmarks were based mainly on data from dominantly male Caucasian cohorts in the Global North, potentially limiting their universality and relevance across diverse demographic groups. Moreover, the conventional methodology of assessing systolic (SBP) and diastolic blood pressure (DBP) as separate quantities fails to consider the clinical significance of their interdependence. The current standardized BP thresholds, while helpful in prescreening potential hypertensive individuals, lack precision as biomarkers for categorizing the severity of hypertension or as accurate quantitative indicators of CVD risk.
Therefore, understanding the population-wide distributions of BP across demographics is crucial. In this study, we investigate how various demographic factors, such as biological sex, age, and race, impact BP values. We also investigate the cross correlation between SBP and DBP on a population-wide level, and how it varies across demography.
\section{Literature review}
Demographic factors are known to have physiological impacts that influence BP measurements \cite{khajedaluee2016prevalence}. For example, BP values of males typically exhibit higher levels than those of females \cite{Reckelhoff2001-td}. Race and ethnicity are believed to impact the BP \cite{hardy2021racial}; although it is unclear whether BP variations across race/ethnicity is biology-driven (genetic, physiological, or anatomical) or impacted by social determinants of health. Age is another significant demographic factor. As people age, changes in the cardiovascular system often result in fluctuations in BP levels \cite{miall1967relation}. Overall, BP values have been reported to gradually increase with age \cite{carrico2013predictive}.
\section{Method}
\subsection{Dataset}
In this study, we analyzed a very large dataset comprising over 94 million patient encounters containing cuff-based (noninvasive) BP measurements across varied demographics and clinical outcomes. The data were collected at Emory Healthcare (Georgia, USA), between 2000 and 2022. The study has been approved by Emory University's Institutional Review Board under the protocol ID: STUDY00006568. 
\subsection{Pre-processing}
The data was first cleaned, by removing any non-numeric and non-positive BP values, due to human errors. The scatter plots of the SBP and DBP also showed occasional outlier values, which may be due to human error (in entering the data in electronic health records), uncalibrated BP devices, or extreme cases. Over 99\% of SBP and DBP values fell within the ranges of 20 to 200\,mmHg and 30 to 300\,mmHg, respectively. For the current population-wide statistical analysis, we excluded BP values outside these ranges. Due to the long data collection span, there were duplicate BP records across electronic health records (with the same patient and encounter IDs and admission time), and there were records with missing SBP or DBP entries, which we excluded during the data cleaning process. Upon data cleaning, we ended up with 75,636,128 records from 2,054,462 unique patients with valid BP values. For the demographics study, a comprehensive assessment of the demographics was conducted to identify and filter out any missing information, including unknown sex and race. The subjects' ages were calculated using their date of birth (DOB) and the corresponding BP recorded times. It was noticed that in some records, DOB or BP measuring times were inconsistent (e.g., the birth date was later than the BP recording time). These records were excluded. We also filtered out records with ages exceeding 120 years old.

As part of the data cleaning process, we observed an over-prevalence of both SBP and DBP values at multiples of 10\,mmHg. Presumably, the right-side digit of the BP alone does not have any clinical significance (it can be any digit between 0 to 9), and statistically, assuming SBP and DBP were independent random variables (as a simplified assumption) the probability of SBP and DBP, both being a multiple of 10\,mmHg is only 1\% (10\% for SBPs times 10\% for DBPs, which both have 0 as their right digit). However, our data showed a prevalence of 2.86\% for these instances, which was also confirmed from the 2D-histograms of the SBP-DBP values. This implies potentially inaccurate BP self-reports, rounded values, or empirical normal/abnormal BP values guessed by the care-provider without actual measurements--- which is understandable for a very large dataset collected in real clinical settings. Therefore, we excluded all records for which SBP and DBP were both whole multiples of 10\,mmHg from our statistical analysis. We are cognizant that this step will remove around 1\% of the valid and accurate BP measurements as well; but this is in favor of removing the misreported values. 
As a result the cleaned study dataset, comprising all demographic features and BP values, has 69,802,762 records, linked to 1,699,955 unique patients. In this data, 30.49\% of patients have a single record, 18.56\% have two records, 50.95\% have three or more records, and 2.45\% of patients have more than 100 records. For intensive care unit (ICU) patients, due to the sensitivity of patient conditions, BP values are measured regularly and automatically, e.g., every 15-30 minutes. To ensure a fair analysis across patients, we grouped the data based on the median of BP measurements of patients at a given age. This summarized the BP data into 5,317,436 median BP values. 
\subsection{Data distribution}
The dataset consists of 58.9\% females and 41.1\% males.
It encompasses seven race and ethnic groups: Caucasian or White, African American or Black, Asian, Multiple, American Indian or Alaskan Native, Native Hawaiian or Other Pacific Islander, and Hispanic. 
To facilitate age analysis, we categorized individuals into decade-long intervals, with the exception of ages below 20 and above 90, which were consolidated into separate groups. 
\section{Dataset Analysis}
In this section, we present our findings from analyzing the collected dataset. 
\subsection{Impact of Biological Sex on Blood Pressure}
\begin{figure}[tb]
\centering
\includegraphics[width=0.95\columnwidth, trim=0.35cm 3.74cm 0cm 5.6cm, clip]{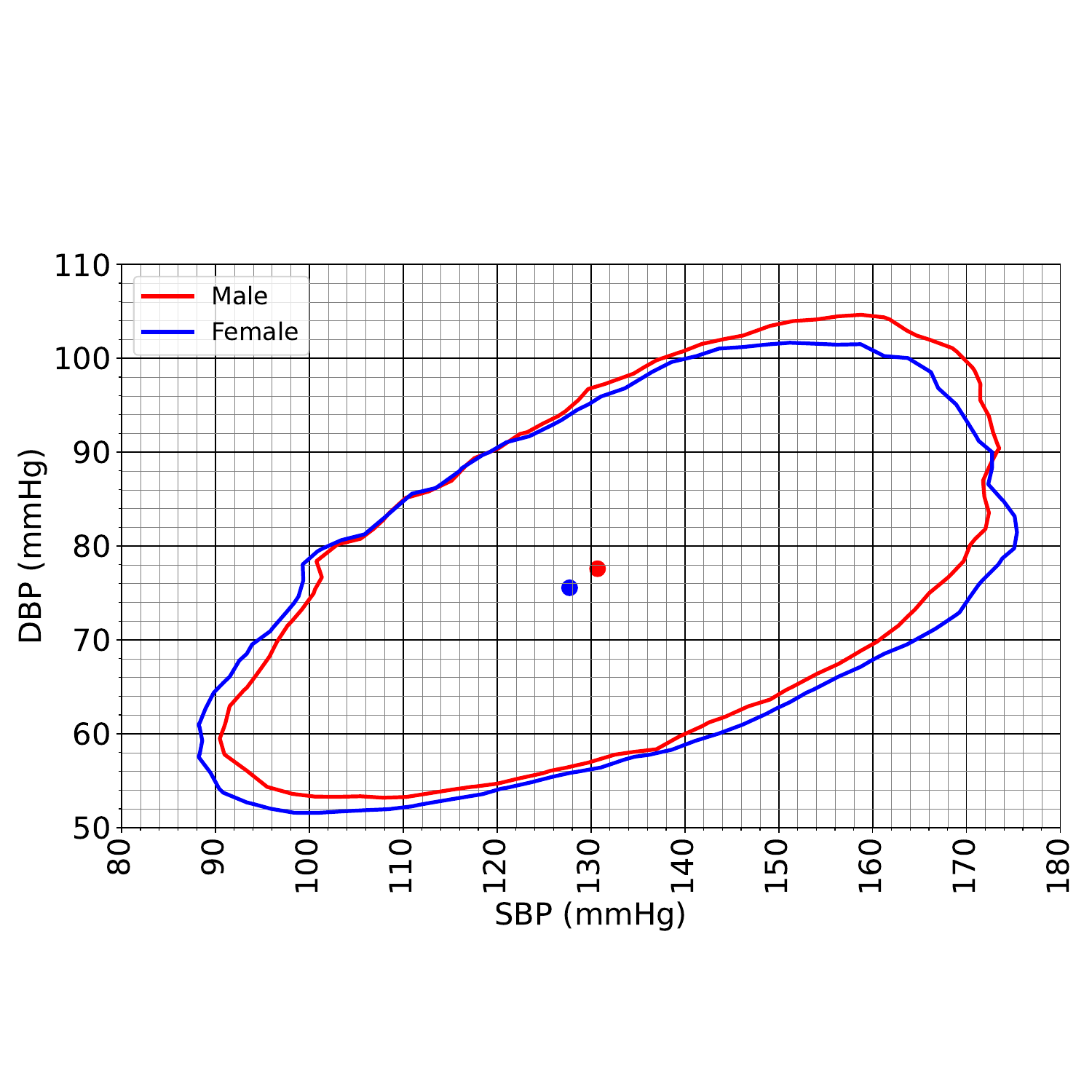}
\caption{Male vs female blood pressure distributions. The contours correspond to 95\% percentile range of the data. Red/blue dots show the mean SBP and DBP of males/females.}\label{Fig: KDE_gender}
\end{figure}

To analyze the impact of sex on BP values, we categorized the final pre-processed dataset into females and male groups. Subsequently, we calculated the mean and standard deviation (STD) of SBP and DBP for each group. Our findings revealed significant differences in mean BP levels between females and males, as summarized in Table~\ref{tab: Gender_statistics}. The difference between the mean SBP and DBP values between the two sex groups are 2.98 and 2.03\,mmHg, respectively. Fig.~\ref{Fig: KDE_gender} illustrates the 95\% percentile range contours for males vs females BP distribution. The mean SBP and DBP for male and female groups are marked with dots.
\begin{table}
    \centering
    \caption{Summary of sex-specific BP statistics, including the number and percentage of BP records, mean and STD, along with the correlation coefficient between SBP and DBP.}
    \label{tab: Gender_statistics}
    \begin{tabular}
    {p{0.9cm} p{.9cm} p{0.3cm}p{0.6cm} p{.6cm}p{.4cm} p{.4cm} p{1.35cm}}
        \hline
        Sex & Number & \% & mean & std & mean & std &  $\rho$  \\
              & &  & SBP & SBP & DBP & DBP  & (SBP, DBP) \\
        \hline
        Female & 3,134,108 & 58.9 & 127.72 & 17.99 & 75.55 & 9.96 & 0.56 \\
        Male & 2,183,328 & 41.1 & 130.70 & 16.70 & 77.58 & 10.23  & 0.60 \\
        \hline
    \end{tabular}
\end{table}
\subsection{Impact of Race/Ethnicity on Blood Pressure Values}
\begin{figure}[tb]
\centering
\includegraphics[width=0.95\columnwidth,trim=0.35cm 1.5cm 0.25cm 3.5cm, clip]{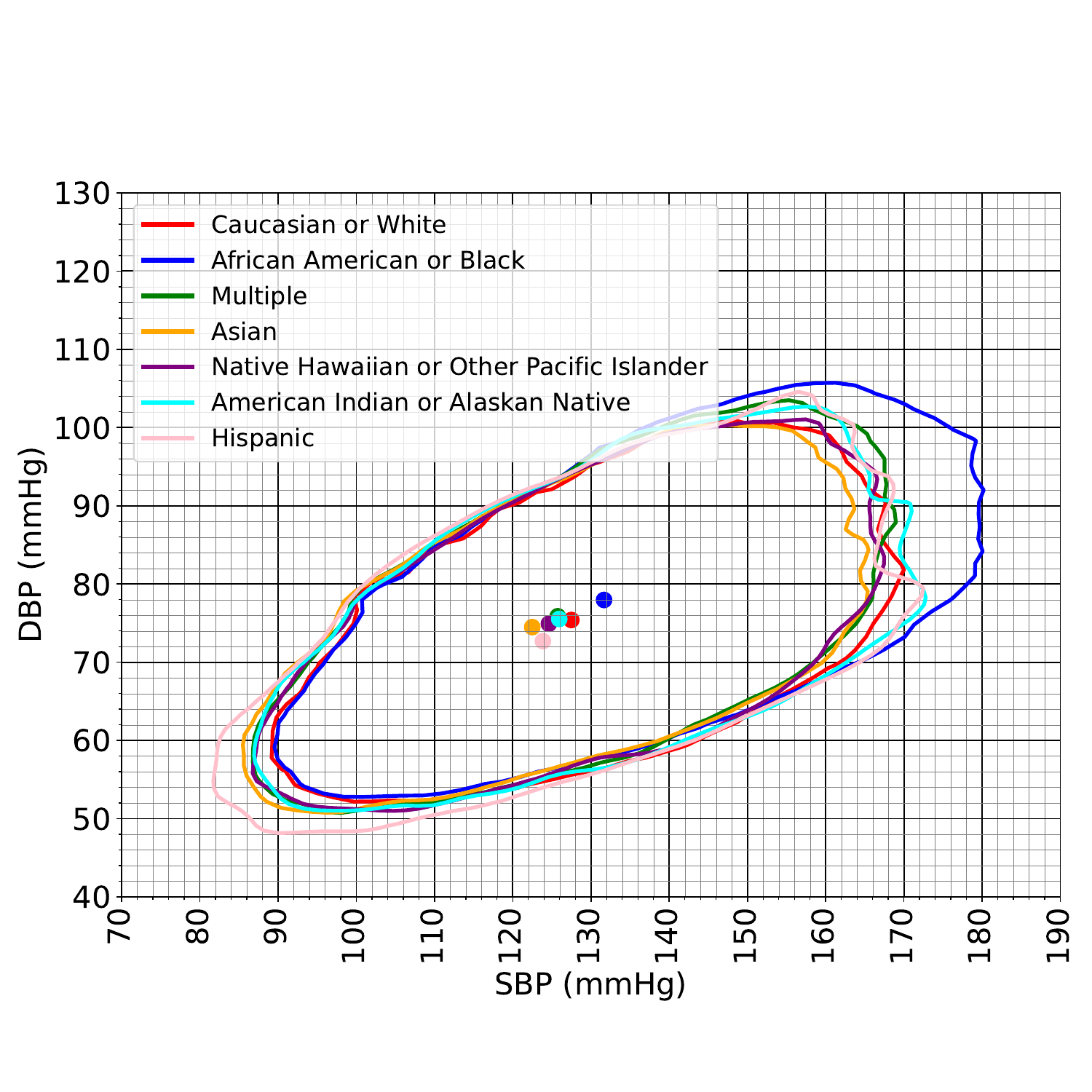}
\caption{Blood pressure distribution across race/ethnicity. The contours correspond to 95\% percentile ranges. Dots show the mean SBP and DBP in each racial/ethnic group.}\label{Fig: KDE_race}
\end{figure}

In exploring the influence of racial and ethnic groups on BP values, we categorized the final pre-processed dataset based on the racial/ethnic groups listed in Table~\ref{tab: Race_statistics}. Importantly, race and ethnicity are compound factors, impacted by biology (genetic factors that impact BP from a physiological or anatomical standpoint) and also by social determinants of health (factors related to race/ethnicity disparities). Table~\ref{tab: Race_statistics} presents the mean and STD for DBP and SBP across racial and ethnic groups. Accordingly, the African American or Black group had the highest mean SBP and DBP values. The Asian population demonstrates the lowest SBP, while the Hispanic group has the minimum DBP. 
In Fig.~\ref{Fig: KDE_race}, the mean and the 95\% percentile range contours of each racial and ethnic group are visualized. Accordingly, the African American or Black group exhibits a distinctive pattern towards higher SBP and DBP, while the Hispanic group tends to have lower SBP and DBP, when contrasted with other groups.
\begin{table*}
    \centering
    \caption{Summary of race and ethnic-specific blood pressure statistics, including the number and percentage of records, mean and standard deviation for the median of SBP and DBP values, along with the SBP-DBP correlation coefficient.}
    \label{tab: Race_statistics}
    \begin{tabular}{rrcccccc}
        \hline
        Racial and ethnic category & Number & \% & mean & std & mean & std &  $\rho$  \\
        &  &  & SBP & SBP & DBP & DBP &  (SBP, DBP)\\
        \hline
         Caucasian or White & 2,983,822 & 56.1 & 127.49 & 16.36 & 75.40 & 9.66 & 0.53 \\
         
          African American or Black & 2,089,345 & 39.3 & 131.67  & 18.75 & 77.96 & 10.58  & 0.62 \\
          
        Asian & 172,233 & 3.2 & 122.49 & 16.98 & 74.49 & 9.67 & 0.59 \\
        
        Multiple & 40,520 & 0.8 & 125.75 & 17.06 & 75.88 & 10.09 & 0.62 \\
        
        American Indian or Alaskan Native & 14,779 & 0.3 &  125.94 & 17.45 & 75.53 & 9.98 & 0.59 \\
        
        Native Hawaiian or Other Pacific Islander & 11,632 & 0.2 & 124.62 & 16.50 & 74.92 & 9.88 & 0.61 \\
        
Hispanic & 5,105 & 0.1 & 123.85 & 17.81 & 72.70 & 10.52  & 0.63\\
        \hline
    \end{tabular}
\end{table*}
\subsection{Impact of Age on Blood Pressure Values}
\begin{figure}[tb]
\centering
\includegraphics[width=0.91\columnwidth, trim=0.35cm 0.41cm 0cm 0.4cm, clip]{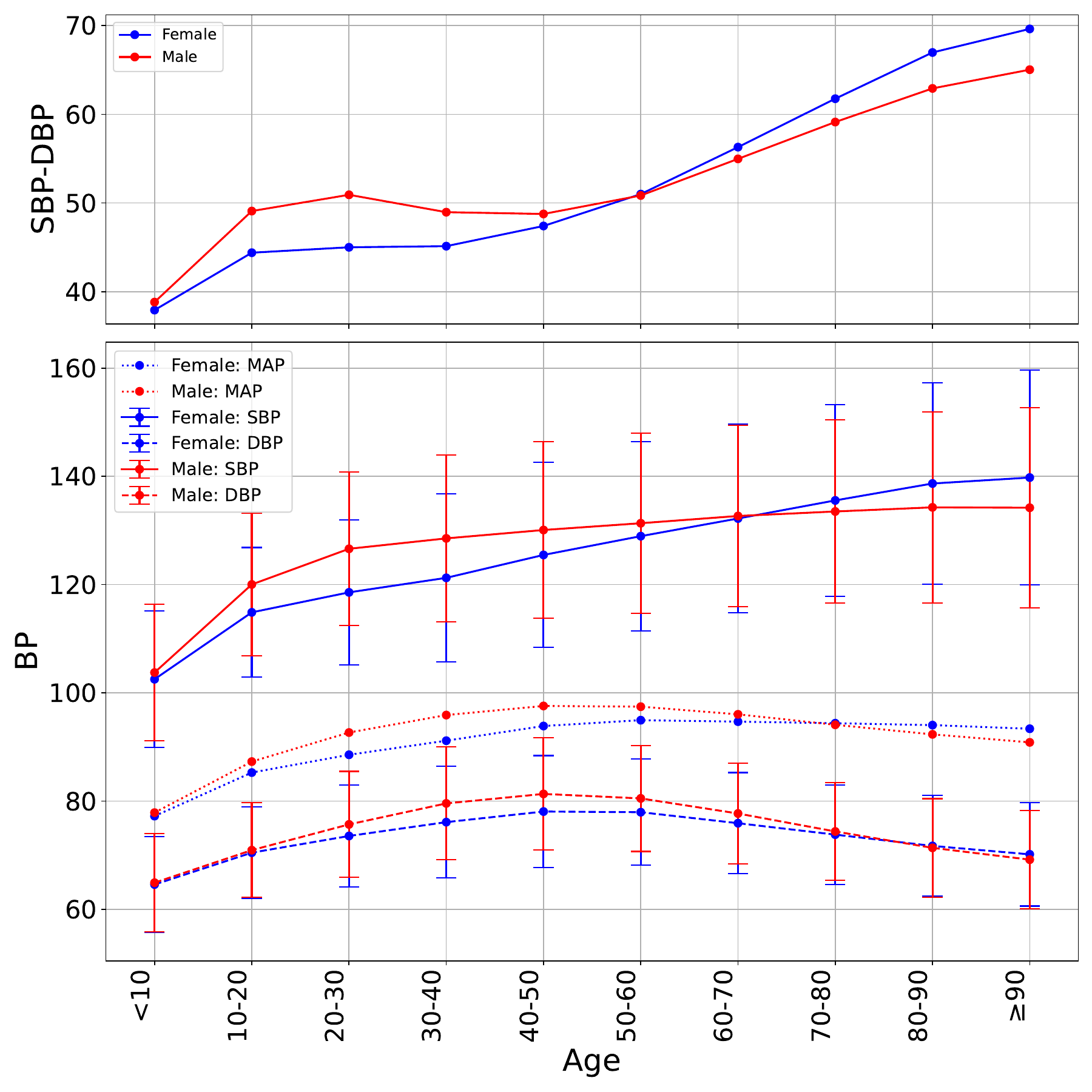}
\caption{Mean$\pm$STD pulse pressure (top) and blood pressure (bottom) by age, for males and females.}\label{Fig: SBP_DBP_age_Bar}
\end{figure}

Table~\ref{tab: Age_statistics} provides a summary of the BP statistics across different age groups. The maximum difference in mean SBP among age groups is 23.06\,mmHg, with the age group of 90 and above exhibiting the highest mean SBP, while the age group less than 20 shows the minimum mean SBP. Similarly, for mean DBP, the maximum difference is 9.56\,mmHg, observed in the 40--49 age group as the maximum mean DBP, whereas the age groups of 90 and above and less than 20 demonstrate the lowest DBP.
\begin{table}
    \centering
    \caption{Age-specific blood pressure statistics, including the number and percentage of records, mean and STD of SBP and DBP, and their correlation coefficient.}
    \label{tab: Age_statistics}
    \begin{tabular}
    {p{0.68cm} p{0.9cm} p{0.3cm}p{0.6cm} p{.5cm}p{.5cm} p{.5cm} p{1.4cm}}
        \hline
        Age & Number & \% & mean & std & mean & std &  $\rho$ \\
            &&     & SBP & SBP & DBP & DBP  & (SBP, DBP)\\
        \hline
$<$ 20 & 193,631 & 3.6 &115.05 & 13.65 & 69.80 & 8.90 & 0.61  \\
           20-29 & 547,023 & 10.3 & 121.27 & 14.19 & 74.27 & 9.60 & 0.66 \\ 
           30-39 & 674,798 & 12.7 & 123.79 & 15.91 & 77.31 & 10.48 & 0.74\\          
           40-49 & 806,952 & 15.2 & 127.32 & 16.93 & 79.36 & 10.46 & 0.73\\      
         50-59 & 984,094 & 18.5 & 129.98 & 17.19 & 79.05 & 9.88 & 0.67 \\  
        60-69 & 1,012,979 & 19.1 & 132.42 & 17.13 & 76.71 & 9.38 & 0.58 \\       
        70-79 & 732,165 & 13.8 & 134.62 & 17.36 & 74.06 & 9.11 & 0.52 \\ 
        
        80-89 & 305,157 & 5.7 & 136.89 & 18.35 & 71.56 & 9.22 & 0.49\\ 
        $\geq$ 90 & 60,637 & 1.1 & 138.11 & 19.59 & 69.86 & 9.44 & 0.50 \\ 
        \hline
    \end{tabular}
\end{table}
Fig.~\ref{Fig: SBP_DBP_age_Bar} shows the mean BP within each age decade. The average SBP values demonstrate an increasing trend with age, ranging from approximately 100\,mmHg in children to 130--140\,mmHg in the elderly. While the average DBP across all age groups is within the range of approximately 60 to 80\,mmHg, with a hump in the 40--50 age group. Notably, the average SBP and DBP values of males generally exceeds those of females from adolescence up to the 60--70 age range, after which the average SBP of females exceeds the SBP of males. The male SBP almost plateaus at 60 years of age and above.

\begin{figure}[tb]
\centering
\includegraphics[width=0.9\columnwidth,trim=0.40cm 3.4cm 0cm 5.7cm, clip]{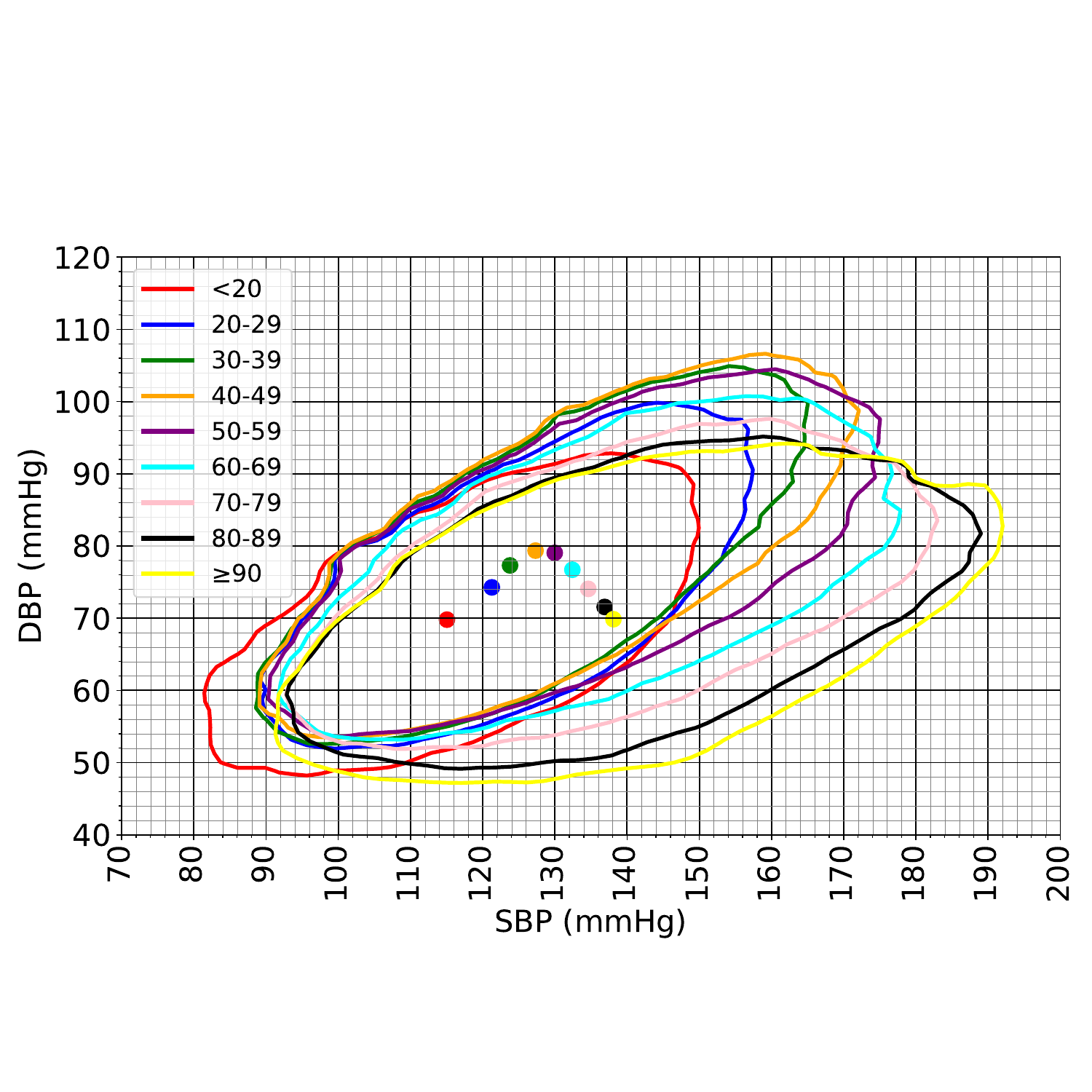}
\caption{Blood pressure distribution of different age groups. The contours correspond to 95\% percentile ranges. Dots show the mean  SBP and DBP in each age group.}\label{Fig: KDE_age}
\end{figure}

The Mean Arterial Pressure (MAP) follows a similar trend as the DBP, but after the age of 70, it plateaus at around 90\,mmHg. Fig.~\ref{Fig: SBP_DBP_age_Bar}, also shows the pulse pressure (PP), which is the difference between SBP and DBP. Accordingly, up to 50 years of age, the average male PP exceeds the average female PP, but the trend changes after 50 years of age, and the average female PP exceeds the average male PP. These results exactly align with a recent research by Chou et al., based on a study involving 162,636 participants over two decades~\cite{Chou2022}. Fig. \ref{Fig: KDE_age} illustrates the mean and 95\% percentile range contours associated with each age group, presenting a distinct pattern across different age groups. To note, besides the average SBP and DBP patterns, the standard deviations and correlation coefficients between SBP and DBP also change with age.
\section{Discussion}
BP is impacted by demographic factors. Both age and sex play a significant role in this regard. BP also varies across race and ethnicity, presumably, due to both biological factors and social determinants of health. The BP literature typically provides mean and STD values for SBP and DBP across various demographic features. We complemented this by studying the correlation coefficients between SBP and DBP. Males typically show a stronger correlation coefficient compared to females, alongside their elevated mean SBP and DBP values. Moreover, the Hispanic group displays the highest correlation coefficient despite having lower mean DBP values. Furthermore, the 30-39 and 40-49 age groups demonstrate the highest correlation coefficient. This underscores the significance of correlation coefficients in offering insights beyond mean and STD values. 
\section{Conclusion}
The study highlighted the significance of considering demographic factors in BP analysis and highlighted the variability of BP across sex, age, and race/ethnicity. In this work, we followed a data-driven approach, examining the existing BP data alone. This information can be used in the future to model the relationship between different demographic factors and BP, aiding in the development of machine learning algorithms for personalized BP assessment and hypertension risk assessment.
\bibliographystyle{IEEEtran}
\bibliography{References}
\end{document}